\documentclass{ws-procs961x669}            
\begin{document}

\def\eqa{\!\!&=&\!\!}
\def\ccr{\nonumber\\}

\def\la{\langle}
\def\ra{\rangle}

\def\del{\Delta}
\def\ddel{{}^\bullet\! \Delta}
\def\deld{\Delta^{\hskip -.5mm \bullet}}
\def\ddeld{{}^{\bullet}\! \Delta^{\hskip -.5mm \bullet}}
\def\dddel{{}^{\bullet \bullet} \! \Delta}

\newcommand{\ba}{\begin{array}}
\newcommand{\ea}{\end{array}}
\newcommand{\identy}{1\!\!1}
\def\ni{\noindent}
\def\la{\langle}
\def\ra{\rangle}

\def\rld{\rlap{\,/}D}
\def\rldd{\rlap{\,/}\nabla}
%
\def\half{{1\over 2}}
\def\third{{1\over3}}
\def\fourth{{1\over4}}
\def\fifth{{1\over5}}
\def\sixth{{1\over6}}
\def\seventh{{1\over7}}
\def\eigth{{1\over8}}
\def\ninth{{1\over9}}
\def\tenth{{1\over10}}
\def\bN{\mathop{\bf N}}
\def\R{{\rm I\!R}}
\def\Eins{{\mathchoice {\rm 1\mskip-4mu l} {\rm 1\mskip-4mu l}
{\rm 1\mskip-4.5mu l} {\rm 1\mskip-5mu l}}}
\def\Z{{\mathchoice {\hbox{$\sf\textstyle Z\kern-0.4em Z$}}
{\hbox{$\sf\textstyle Z\kern-0.4em Z$}}
{\hbox{$\sf\scriptstyle Z\kern-0.3em Z$}}
{\hbox{$\sf\scriptscriptstyle Z\kern-0.2em Z$}}}}
\def\abs#1{\left| #1\right|}
\def\com#1#2{
        \left[#1, #2\right]}
\def\square{\kern1pt\vbox{\hrule height 1.2pt\hbox{\vrule width 1.2pt
   \hskip 3pt\vbox{\vskip 6pt}\hskip 3pt\vrule width 0.6pt}
   \hrule height 0.6pt}\kern1pt}
      \def\boxop{{\raise-.25ex\hbox{\square}}}
\def\contract{\makebox[1.2em][c]{
        \mbox{\rule{.6em}{.01truein}\rule{.01truein}{.6em}}}}
\def\ltap{\ \raisebox{-.4ex}{\rlap{$\sim$}} \raisebox{.4ex}{$<$}\ }
\def\gtap{\ \raisebox{-.4ex}{\rlap{$\sim$}} \raisebox{.4ex}{$>$}\ }
\def\mn{{\mu\nu}}
\def\rs{{\rho\sigma}}
\newcommand{\Det}{{\rm Det}}
\def\Tr{{\rm Tr}\,}
\def\tr{{\rm tr}\,}
\def\sumij{\sum_{i<j}}
\def\e{\,{\rm e}}
\def\pa{\partial}
\def\dA{\partial^2}
\def\ddx{{d\over dx}}
\def\ddt{{d\over dt}}
\def\der#1#2{{d #1\over d#2}}
\def\lie{\hbox{\it \$}} 
\def\partder#1#2{{\partial #1\over\partial #2}}
\def\secder#1#2#3{{\partial^2 #1\over\partial #2 \partial #3}}
%
\newcommand{\be}{\begin{equation}}
\newcommand{\ee}{\end{equation}\noindent}
\newcommand{\bear}{\begin{eqnarray}}
\newcommand{\ear}{\end{eqnarray}\noindent}
\newcommand{\benn}{\begin{enumerate}}
\newcommand{\enn}{\end{enumerate}}
\newcommand{\veject}{\vfill\eject}
\newcommand{\ven}{\vfill\eject\noindent}
%
\def\eq#1{{eq. (\ref{#1})}}
\def\eqs#1#2{{eqs. (\ref{#1}) -- (\ref{#2})}}
%
\def\totint{\int_{-\infty}^{\infty}}
\def\posint{\int_0^{\infty}}
\def\negint{\int_{-\infty}^0}
\def\pint{{\dps\int}{dp_i\over {(2\pi)}^d}}
%
\newcommand{\GeV}{\mbox{GeV}}
\def\FFdual{F\cdot\tilde F}
\def\bra#1{\langle #1 |}
\def\ket#1{| #1 \rangle}
\def\braket#1#2{\langle {#1} \mid {#2} \rangle}
\def\vev#1{\langle #1 \rangle}
\def\rightvac{\mid 0\rangle}
\def\leftvac{\langle 0\mid}
\def\ihbar{{i\over\hbar}}
\def\slash#1{#1\!\!\!\raise.15ex\hbox {/}}
\newcommand{\slD}{\,\raise.15ex\hbox{$/$}\kern-.27em\hbox{$\!\!\!D$}}
\newcommand{\slpartial}{\raise.15ex\hbox{$/$}\kern-.57em\hbox{$\partial$}}
\newcommand{\cL}{\cal L}
\newcommand{\D}{\cal D}
\newcommand{\Dhalf}{{D\over 2}}
\def\eps{\epsilon}
\def\epshalf{{\epsilon\over 2}}
\def\lag{( -\partial^2 + V)}
\def\freeexp{{\rm e}^{-\int_0^Td\tau {1\over 4}\dot x^2}}
\def\kinb{{1\over 4}\dot x^2}
\def\kinf{{1\over 2}\psi\dot\psi}
\def\expk{{\rm exp}\biggl[\,\sum_{i<j=1}^4 G_{Bij}k_i\cdot k_j\biggr]}
\def\expp{{\rm exp}\biggl[\,\sum_{i<j=1}^4 G_{Bij}p_i\cdot p_j\biggr]}
\def\expshort{{\e}^{\half G_{Bij}k_i\cdot k_j}}
\def\expabb{{\e}^{(\cdot )}}
\def\epseps#1#2{\varepsilon_{#1}\cdot \varepsilon_{#2}}
\def\epsk#1#2{\varepsilon_{#1}\cdot k_{#2}}
\def\kk#1#2{k_{#1}\cdot k_{#2}}
\def\G#1#2{G_{B#1#2}}
\def\Gp#1#2{{\dot G_{B#1#2}}}
\def\GF#1#2{G_{F#1#2}}
\def\Dab{{(x_a-x_b)}}
\def\Dsq{{({(x_a-x_b)}^2)}}
\def\PITD{{(4\pi T)}^{-{D\over 2}}}
\def\4piTD{{(4\pi T)}^{-{D\over 2}}}
\def\4piT4{{(4\pi T)}^{-2}}
\def\TintmD{{\dps\int_{0}^{\infty}}{dT\over T}\,e^{-m^2T}
    {(4\pi T)}^{-{D\over 2}}}
\def\Tintm4{{\dps\int_{0}^{\infty}}{dT\over T}\,e^{-m^2T}
    {(4\pi T)}^{-2}}
\def\Tintm{{\dps\int_{0}^{\infty}}{dT\over T}\,e^{-m^2T}}
\def\Tint{{\dps\int_{0}^{\infty}}{dT\over T}}
\def\np{n_{+}}
\def\nm{n_{-}}
\def\Np{N_{+}}
\def\Nm{N_{-}}
\newcommand{\slG}{{{\dot G}\!\!\!\! \raise.15ex\hbox {/}}}
\newcommand{\Gd}{{\dot G}}
\newcommand{\Gund}{{\underline{\dot G}}}
\newcommand{\Gdd}{{\ddot G}}
\def\GBd12{{\dot G}_{B12}}
\def\Dx{\dps\int{\cal D}x}
\def\Dy{\dps\int{\cal D}y}
\def\Dpsi{\dps\int{\cal D}\psi}
\def\dint#1{\int\!\!\!\!\!\int\limits_{\!\!#1}}
\def\ddtau{{d\over d\tau}}
\def\ie{\hbox{$\textstyle{\int_1}$}}
\def\iz{\hbox{$\textstyle{\int_2}$}}
\def\id{\hbox{$\textstyle{\int_3}$}}
\def\ldop{\hbox{$\lbrace\mskip -4.5mu\mid$}}
\def\rdop{\hbox{$\mid\mskip -4.3mu\rbrace$}}
%
\newcommand{\1}{{\'\i}}
\newcommand{\no}{\noindent}
\def\non{\nonumber}
\def\dps{\displaystyle}
\def\sy{\scriptscriptstyle}
\def\sy{\scriptscriptstyle}

%

\newcommand{\bea}{\begin{eqnarray}}  
\newcommand{\eea}{\end{eqnarray}}  
\def\eqa{&=&}  
\def\ccr{\nonumber\\}  
  
\def\a{\alpha}
\def\b{\beta}
\def\m{\mu}
\def\n{\nu}
\def\r{\rho}
\def\s{\sigma}
\def\ep{\epsilon}

\def\cosech{\rm cosech}
\def\sech{\rm sech}
\def\coth{\rm coth}
\def\tanh{\rm tanh}

\title{Tadpole contribution to magnetic photon-graviton conversion}

\author{N. Ahmadiniaz}

\address{
Helmholtz-Zentrum Dresden-Rossendorf, Bautzner Landstra\ss e 400, 01328 Dresden, Germany\\
E-mail: n.ahmadiniaz@hzdr.de}

\author{F. Bastianelli}

\address{Dipartimento di Fisica ed Astronomia, Universit\`a di Bologna, Via Irnerio 46, I-40126 Bologna, Italy
and INFN, Sezione di Bologna, Via Irnerio 46, I-40126 Bologna, Italy\\
E-mail: bastianelli@bo.infn.it}

\author{F. Karbstein}

\address{Helmholtz-Institut Jena, Fr\"obelstieg 3, 07743 Jena, Germany and Theoretisch-Physikalisches Institut, Friedrich-Schiller-Universit\"at Jena, Max-Wien-Platz 1, 07743 Jena, Germany\\
E-mail: felix.karbstein@uni-jena.de}

\author{{\underline {C. Schubert}}} 

\address{Instituto de F\'isica y Matem\'aticas
Universidad Michoacana de San Nicol\'as de Hidalgo
Edificio C-3, Apdo. Postal 2-82
C.P. 58040, Morelia, Michoac\'an, M\'exico\\
E-mail: christianschubert137@gmail.com}

\begin{abstract}
Photon-graviton conversion in a magnetic field is a process that is usually studied at tree level, but the one-loop corrections due to scalars and spinors have also been calculated. Differently from the tree-level process, at one-loop one finds the amplitude to depend on the photon polarization, leading to dichroism. However, previous calculations overlooked a tadpole contribution of the type that was considered to be vanishing in QED for decades but erroneously so, as shown by H. Gies and one of the authors in 2016. Here we compute this missing diagram in closed form, and show that it does not contribute to dichroism. 
\end{abstract}


\bodymatter

\section{Introduction: photon-graviton conversion}\label{aba:sec1}

Einstein-Maxwell theory contains a tree-level vertex for 
photon-graviton conversion in a constant electromagnetic field:
\bear
\half
\kappa 
h_{\mn}\Bigl(F^{\mu\alpha}f^{\nu}_{\,\,\,\alpha} + f^{\mu}_{\,\,\alpha}\,F^{\nu\alpha}
\Bigr) - \fourth\kappa h^{\mu}_{\mu}F^{\alpha\beta} f_{\alpha\beta}.
\ear
Here $h_{\mu\nu}$  denotes the graviton, $f_{\mu\nu}$ 
 the photon, $F^{\mu\nu}$  the external field, and
$\kappa$  the gravitational coupling constant.

This interaction leads to photon-graviton oscillations similar to the better-known neutrino or
photon-axion oscillations \cite{gertsenshtein,zelnovbook,rafsto,sikivie,morris,losotr,magueijo,chen,cilhar,ardidv,defuza}
(see \cite{eecpg} for a recent application to gravitational waves). 

In momentum space, this vertex becomes 
\bear
\hspace{-30pt}
\Gamma^{\rm (tree)}(k,\varepsilon;F) &=& \epsilon_{\mn}\varepsilon_{\alpha} \Pi^{\mn,\alpha}_{\rm (tree)} (k;F)  
, \quad
 \Pi^{\mn,\alpha}_{\rm (tree)} (k;F)  = -{i \kappa\over 2} C^{\mn,\alpha}
 \label{phogravtree}
\ear
with 
\be
C^{\mn,\alpha} =
F^{\mu\alpha}k^{\nu} +F^{\nu\alpha}k^{\mu}
-\bigl(F\cdot k\bigr)^{\mu}\delta^{\nu\alpha}
-\bigl(F\cdot k\bigr)^{\nu}\delta^{\mu\alpha} 
+ \bigl(F\cdot k\bigr)^{\alpha}\delta^{\mn} 
\, .
\label{defCmna}
\ee
Since in a constant field the four-momentum is preserved, $k^\mu$ here is the four-momentum
of the photon as well as of the graviton. 

For the photon polarizations, as is customary we will use the Lorentz frame where
${\bf E}$ and $\bf B$ are collinear, and choose the polarization basis
$\varepsilon_{\perp}^{\alpha}, \varepsilon_{\parallel}^{\alpha}$, where
$\varepsilon_{\parallel}^{\alpha}$ lies in the plane spanned by $k$ and 
$\bf B$ or $\bf E$, and $\varepsilon_{\perp}^{\alpha}$ is perpendicular to it. 
For convenience we construct also the graviton polarization tensor
using the same vectors:
\bear
\varepsilon^{\oplus\mu\nu} = \varepsilon^{\perp\mu}\varepsilon^{\perp\nu}
- \varepsilon^{\parallel\mu}\varepsilon^{\parallel\nu}, \quad
\varepsilon^{{\otimes}\,\mu\nu} = \varepsilon^{\perp\mu}\varepsilon^{\parallel\nu}
+ \varepsilon^{\parallel\mu}\varepsilon^{\perp\nu} \, .
\ear
From the CP-invariance of Einstein-Maxwell theory one can then derive the following
selection rules:
\begin{itemize}

\item
For a  purely magnetic field  $\varepsilon^{\oplus}$  couples only
to  $\varepsilon^{\perp}$   and  
 $\varepsilon^{\otimes}$  only
to  $\varepsilon^{\parallel}$. 

\item
For a  purely electric field  $\varepsilon^{\oplus}$  couples only
to  $\varepsilon^{\parallel}$  and  
 $\varepsilon^{\otimes}$  only
to  $\varepsilon^{\perp}$.

\end{itemize}

\section{One-loop photon-graviton vacuum polarization}

\begin{figure}
{\centering
\hspace{100pt}
\includegraphics[width=1.5in]{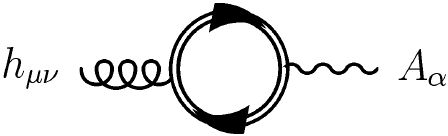}
}
\caption{One-loop correction to the photon-graviton amplitude in a constant field.}
\label{fig-phograv}
\end{figure}

In \cite{61,71} the worldline formalism was used along the lines of
\cite{strassler1,rss,vv} to study the one-loop corrections to this amplitude 
due to a scalar or spinor loop, see Fig. \ref{fig-phograv}. 

\begin{figure}
{\centering
\hspace{50pt}
\includegraphics[width=3.5in]{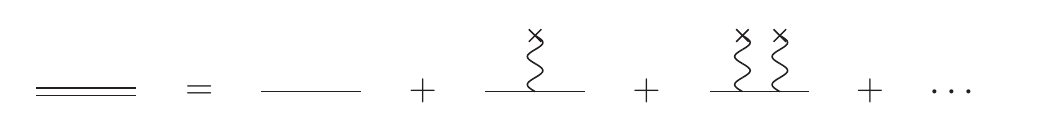}
}
\caption{Full scalar or spinor propagator in a constant field.}
\label{fig-fullprop}
\end{figure}

Here we employ the usual double-line notation for the full propagator in the 
external electromagnetic field, Fig. 
\ref{fig-fullprop}. 

These calculations lead to the same type of two-parameter integrals as for the
well-studied case of the one-loop photon vacuum polarisation in a constant field 
\cite{toll,minguzzi,baibre,biabia,adler,batsha,ritus_annphys,tsaerb,bakast,melsto,ditgiebook,kohyam}. 
Both the scalar and the spinor-loop amplitudes are UV divergent, but multiplicatively renormalizable.
For example, the scalar-loop contribution in dimensional regularization displays the pole 
\bear
\Pi_{\rm scal, div}^{\mn,\alpha}(k)
&=&
{ie^2\kappa\over 3(4\pi)^2}{1\over D-4}C^{\mn,\alpha}
\nonumber
\ear

\no
where $
C^{\mn,\alpha}$
is the tree level vertex \eqref{defCmna}.

For studying the relative importance of the one-loop amplitudes
it is useful to normalize them by the tree-level amplitude
( the `bar' on $\Pi$ denotes renormalization)

\bear
\hat \Pi^{Aa}_{\rm scal,spin}
(\hat\omega,\hat B,\hat E) &\equiv& 
{\bar \Pi^{Aa}_{\rm scal, spin}(\hat\omega,\hat B,\hat E)
\over -{i\over 2}\kappa C^{Aa}}
\ear
\medskip
\no
where $A=\oplus,\otimes$  and $a= \perp, \parallel$ ,
$\hat\omega = {\omega\over m}$ ,
$\hat B = {eB\over m^2}$ , 
$\hat E = {eE\over m^2}$.

It then becomes obvious that the one-loop amplitudes become quantitatively relevant 
only for field strengths close to the critical ones $\hat B, \hat E \approx 1$. 
Macroscopic fields of this magnitude are known to exist only for the magnetic case,
so that we will set $\hat E =0$ in the following. In \cite{71} it was shown that 
for this purely magnetic case the parameter integrals allow for a straightforward
numerical evaluation as long as the photon energy $\hat\omega$ is below the pair-creation 
threshold $\hat\omega_{\rm cr}$, 
\bear
\hat\omega_{\rm cr,scal}^{\oplus \perp} =
\hat\omega_{\rm cr,scal}^{\otimes \parallel}  &=& 2\sqrt{1+\hat B}  \, ,
\non\\
\hat\omega_{\rm cr,spin}^{\oplus\perp} &=& 1+\sqrt{1+2\hat B} \, ,
\non\\
\hat\omega_{\rm cr,spin}^{\otimes\parallel} &=& 2 \, .\non\\
\label{omegacrit}
\ear
There also a number of special cases were studied that allow for more explicit representations:

\begin{enumerate}

\item
The case of small $B$ and arbitrary $\omega$ leads to single - parameter integrals over Airy functions.

\item
In the large $B$ limit one finds a logarithmic growth in the field strength,
\bear
\hat \Pi^{Aa}_{\rm scal}
(\hat\omega,\hat B) &\stackrel{\hat B\to\infty}{\sim}&
-{\alpha\over 12\pi}\ln (\hat B)\, ,
 \\ 
\hat \Pi^{Aa}_{\rm spin} 
(\hat\omega,\hat B) &\stackrel{\hat B\to\infty}{\sim}& 
-{\alpha\over 3\pi}\ln (\hat B)
\, .
\ear

\item
In the limit of vanishing photon energy the amplitudes can be related to the corresponding
one-loop effective Lagrangians:
\bear
\hat\Pi_{\rm scal,spin}^{\oplus \perp}(\hat\omega=0,\hat B) 
&=& 
-{2\pi\alpha\over m^4} \Bigl({1\over \hat B}{\pa\over\pa \hat B} + {\pa^2\over \pa \hat B^2}\Bigr)
\,{\cal L}_{\rm scal,spin}^{\rm EH}(\hat B)
\, ,
\label{limitscal}\\
\hat\Pi_{\rm scal,spin}^{\otimes \parallel}(\hat\omega=0,\hat B)
&=& 
-{4\pi \alpha\over m^4}
{1\over \hat B}{\pa\over\pa \hat B}\, {\cal L}_{\rm scal,spin}^{\rm EH}(\hat B)
\, .
\label{limitspin}
\ear
Here ${\cal L}_{\rm scal,spin}^{\rm EH}(\hat B)$ denotes the one-loop
effective Lagrangian in a constant magnetic field, obtained for the spinor QED
case by Heisenberg and Euler \cite{eulhei} and for scalar QED by Weisskopf \cite{weisskopf}:
\bear
{\cal L}_{\rm scal}^{\rm EH}(\hat B)
&=&
-{m^4\over 16\pi^2} 
\int_0^{\infty}
{d\hat s\over \hat s^3}\,\e^{-i\hat s}
\biggl[{\hat B\hat s\over\sin (\hat B\hat s)} - {(\hat B\hat s)^2\over 6} -1 \biggr]
\, ,
\\
{\cal L}_{\rm spin}^{\rm EH}(\hat B)
&=&
\,\,\,\,\,
{m^4\over 8\pi^2} 
\int_0^{\infty}
{d\hat s\over \hat s^3}\,\e^{-i\hat s}
\biggl[{\hat B\hat s\over{\tan} (\hat B\hat s)} + {(\hat B\hat s)^2\over 3} -1 \biggr]
\, .
\label{EHL}
\ear

\end{enumerate}

\section{Dichroism}

For realistic parameters, the one-loop corrections turn out to be small compared to the tree-level
amplitudes. However, there is also a qualitative difference to the
tree-level amplitude. As has been emphasized in \cite{cilhar}, the tree level photon-graviton
conversion does, contrary to the better known photon-axion case, not lead to a dichroism
effect for photon beams. This is because both
photon polarization components have equal conversion rates.
This symmetry gets broken by the loop corrections. Although this 
effect is, of course, tiny and hardly measurable in the near future, 
an exhaustive analysis by Ahlers et al. \cite{ahjari} has shown that it is
still  the leading contribution to magnetic dichroism in the standard model (including standard gravity)!

\section{Quantum electrodynamics in external fields}

\begin{figure}
{\centering
\hspace{130pt}
\includegraphics[width=1.2in]{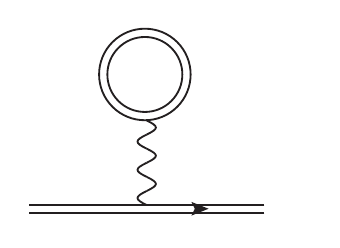}
}
\caption{Full scalar or spinor propagator in a constant field.}
\label{fig-proptadpole}
\end{figure}

In vacuum QED, there is, of course, no one-photon amplitude because of Furry's theorem. 
In the presence of an external field, however, one-photon tadpole diagrams such as the
one in Fig. \ref{fig-proptadpole} will in general be nonzero.

\begin{figure}
{\centering
\hspace{130pt}
\includegraphics[width=.9in]{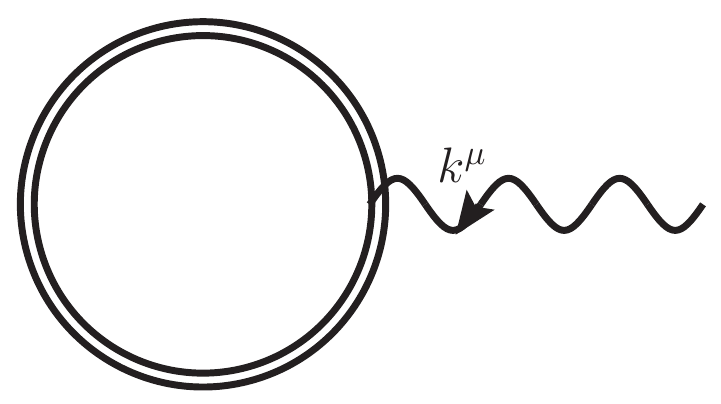}
}
\caption{One photon amplitude in a constant field.}
\label{fig-tadpole}
\end{figure}

If the external field is constant, then the one-photon amplitude Fig. \ref{fig-tadpole} 
can still be shown to vanish.

The argument goes as follows:

\benn

\item
A constant field emits only photons with zero energy-momentum, thus there is a factor of $\delta(k)$.

\item
Because of gauge invariance, this diagram in a momentum expansion starts with the term linear in momentum.

\item
$ \delta (k) k^{\mu} =0$.

\enn

Since the tadpole vanishes, it has been assumed for decades that also any diagram containing it can be
discarded. For example, in the book ``Quantum Electrodynamics with Unstable Vacuum'' by
E.S. Fradkin, D.M. Gitman and S.M. Shvartsman \cite{frgish-book} it is stated (on page 225) even more generally that,
``{\it Thus, in the constant and homogeneous external field combined with that of a plane-wave, all the diagrams containing
the causal current $ {\cal J}^{\mu}(x)$ (i.e., those containing tadpoles having a causal propagator $ S^c(x,y)$),
are equal to zero.}'' 

\begin{figure}
{\centering
\hspace{130pt}
\includegraphics[width=.9in]{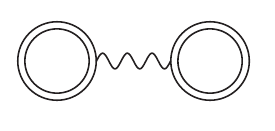}
}
\caption{Handcuff diagram in a constant field.}
\label{fig-handcuff}
\end{figure}

Sometimes also additional arguments have been given; 
in the book 
``Effective Lagrangians in Quantum Electrodynamics'' by Dittrich and Reuter \cite{ditreu-book}
it is argued that the ``handcuff'' diagram of Fig. \ref{fig-handcuff} vanishes because of Lorenz invariance.

However, in 2016 H. Gies and one of the authors \cite{giekar} noted that
such diagrams can give finite values because of the infrared divergence of the connecting photon propagator.
In dimensional regularization, the key integral is
\bear
 \int d^Dk \, \delta^D(k) \frac{k^{\mu}k^{\nu}}{k^2} = \frac{\eta^\mn}{D} \, .
 \label{intk}
\ear
Applying this integral to the handcuff diagram one finds a non-vanishing result, 
which can be expressed in the following simple way in terms of the one-loop
Euler-Heisenberg Lagrangian \cite{karbstein}
\bear
{\cal L}_{\rm spin}^{\rm 1PR}  
&=& \half \partder{{\cal L}_{\rm spin}^{(EH)}}{F^\mn} \partder{{\cal L}_{\rm spin}^{(EH)}}{F_\mn} 
\nonumber
\ear
(the superscript `1PR' stands for ``one-particle reducible''). 
This adds on to the standard diagram for the two-loop Euler-Heisenberg Lagrangian, studied by
V.I. Ritus \cite{ritusspin} half a century ago:

\begin{equation}
{\cal L}_{\rm spin}^{(EH)2-{\rm loop}} = \quad
\raisebox{-3ex}
{\includegraphics[width=1.3in]{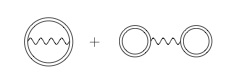}}
\end{equation}

Along the same lines, it was found in \cite{112,113} that the one-loop tadpole contribution Fig.
\ref{fig-proptadpole} to the scalar or spinor propagator in a constant field is also non-vanishing, and
given by
\begin{equation}
S^{\rm 1PR}(p)  = 	\frac{\partial S(p)}{\partial F_{\mu\nu}}\frac{\partial \mathcal{L}^{(EH)}}{\partial F^{\mu\nu}}  
	\nonumber
\end{equation}
where $S(p)$ denotes the tree-level propagator in the field.

\section{Tadpole contribution to the photon-graviton amplitude}

\begin{figure}
{\centering
\hspace{120pt}
\includegraphics[width=.9in]{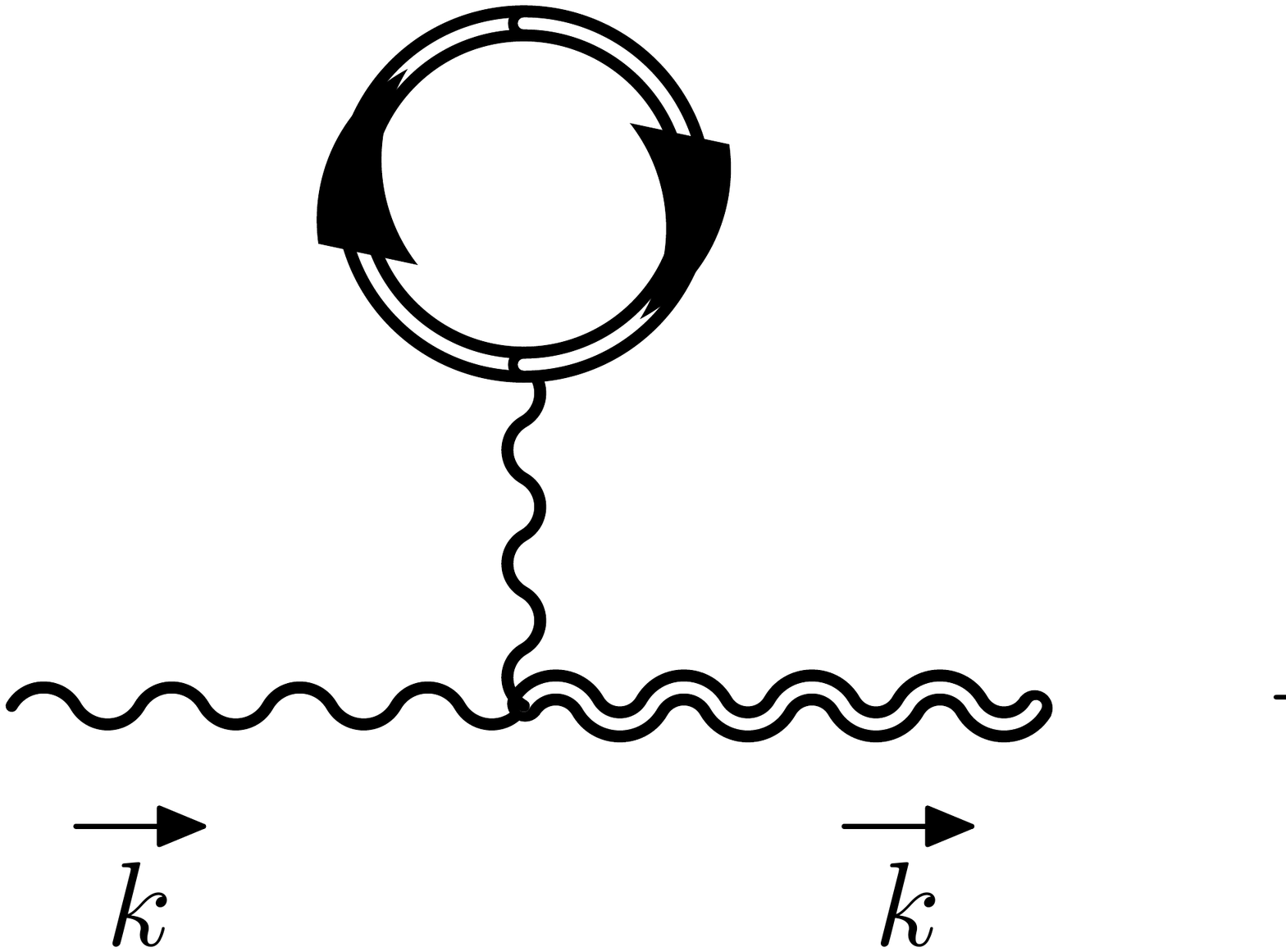}
}
\vspace{-20pt}
\caption{Tadpole contribution to photon-graviton conversion.}
\label{fig-phogravtadpole}
\end{figure}

Returning to the photon-graviton amplitude in a constant field, the point of the present talk is that 
this amplitude, too, has a previously overlooked tadpole contribution, shown in Fig. \ref{fig-phogravtadpole}.

\vspace{-30pt}

\vspace{-10pt}

\begin{figure}
{\centering
\hspace{120pt}
\includegraphics[width=4.2in]{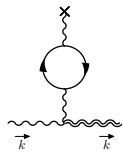}
}
\vspace{-170pt}
\caption{Expanding the tadpole to lowest order in the field.}
\label{fig-tadpolelowest}
\end{figure}

Using the integral \eqref{intk} it is easy to show that its contribution to the 
one-loop amplitudes with a scalar or spinor loop can be written as
\bear
\Gamma^{\rm (tadpole)}_{\rm scal}(k^{\alpha},\varepsilon_\beta;\epsilon_\mn;F_{\kappa\lambda}) &=& 
-i \frac{\alpha}{8\pi} \kappa 
\Bigl(\varepsilon\cdot F \cdot \epsilon \cdot k + \varepsilon\cdot \epsilon\cdot F \cdot k \Bigr)
\nonumber\\
&&\hspace{-50pt} \times
\int_0^{\infty} \frac{dz}{z}
\e^{-\frac{m^2}{eB} z}
\,
 \frac{\coth(z) -1/z}{\sinh(z)}
 \, ,
\\
\Gamma^{\rm (tadpole)}_{\rm spin}(k^{\alpha},\varepsilon_\beta;\epsilon_\mn;F_{\kappa\lambda}) &=& 
i \frac{\alpha}{4\pi} \kappa 
\Bigl(\varepsilon\cdot F \cdot \epsilon \cdot k + \varepsilon\cdot \epsilon\cdot F \cdot k \Bigr)
\nonumber\\
&&\hspace{-50pt} \times
\int_0^{\infty} \frac{dz}{z}
\e^{-\frac{m^2}{eB} z}
\,
 \frac{\coth(z) - \tanh(z) -1/z}{\tanh(z)}
 \, .
\ear
Expanding out the tadpole in powers of the external field we see that the leading term, which is linear
in the field (Fig. \ref{fig-tadpolelowest}), is removed by the photon wave function renormalization.

\vspace{-170pt}

\vspace{10pt}

This gives the renormalized amplitudes 

\bear
\Gamma^{\rm (tadpole)}_{\rm scal, ren}(k^{\alpha},\varepsilon_\beta;\epsilon_\mn;F_{\kappa\lambda}) &=& 
-i \frac{\alpha}{8\pi} \kappa 
\Bigl(\varepsilon\cdot F \cdot \epsilon \cdot k + \varepsilon\cdot \epsilon\cdot F \cdot k \Bigr)
\nonumber\\
&&\hspace{-50pt} \times
\int_0^{\infty} \frac{dz}{z}
\e^{-\frac{m^2}{eB} z}
\,
\Biggl\lbrack  \frac{\coth(z) -1/z}{\sinh(z)}  - \frac{1}{3}\biggr\rbrack
\, ,
\nonumber\\
\Gamma^{\rm (tadpole)}_{\rm spin, ren}(k^{\alpha},\varepsilon_\beta;\epsilon_\mn;F_{\kappa\lambda}) &=& 
i \frac{\alpha}{4\pi} \kappa 
\Bigl(\varepsilon\cdot F \cdot \epsilon \cdot k + \varepsilon\cdot \epsilon\cdot F \cdot k \Bigr)
\nonumber\\
&&\hspace{-50pt} \times
\int_0^{\infty} \frac{dz}{z}
\e^{-\frac{m^2}{eB} z}
\,
\Biggl\lbrack
 \frac{\coth(z) - \tanh(z) -1/z}{\tanh(z)}
  + \frac{2}{3} 
\Biggr\rbrack 
\, .
\nonumber
\ear

\section{Comparison with the main diagram}

These amplitudes are of a structure similar to what we got from the main diagram 
Fig. \ref{fig-phograv} in the limit of low photon energy $ \omega$, eqs. \eqref{limitscal}, 
\eqref{limitspin}. 

However, they  do not contribute to
dichroism  since the polarizations are still bound up in the tree-level vertex 
$ (\varepsilon\cdot F \cdot \epsilon \cdot k + \varepsilon\cdot \epsilon\cdot F \cdot k )$.
Thus the above-mentioned analysis of  Ahlers et al. \cite{ahjari} remains unaffected
by the presence of the tadpole diagram.

\section{Summary and Outlook}

\begin{itemize}

\item
We have presented the first example of a non-vanishing diagram in Einstein-Maxwell theory
involving a tadpole in a constant field. 

\item
This diagram does not contribute to magnetic dichroism. 

\item
A more quantitative analysis is in progress. 

\item
In the  ultra strong-field limit, the tadpoles have been shown even to dominate the (multi-loop)
effective action in QED \cite{karbsteinprl}.
It would be interesting to extend this analysis to the Einstein-Maxwell case.

\end{itemize}



\end{document}